\documentclass[]{elsart}
\usepackage{epsfig}
\input axodraw.sty
\newcommand{\gev}{{\mbox{~GeV}}}
\newcommand{\mett}{{\not\!\!E}_{T}}

\def\beq{\begin{equation}}
\def\eeq{\end{equation}}
\def\bea{\begin{eqnarray}}
\def\eea{\end{eqnarray}}

\newcommand{\threegraphs}[3]{%
\unitlength=1in
\begin{picture}(5.4,5)
\put(1.35,0){\epsfig{file=#3.eps, width=2.5in}}
\put(0,2.5){\epsfig{file=#1.eps, width=2.5in}}
\put(2.7,2.5){\epsfig{file=#2.eps, width=2.5in}}
\put(1.35,2.3){(c)}
\put(0,4.8){(a)}
\put(2.7,4.8){(b)}
\end{picture}}

\newcommand{\twographs}[2]{%
\unitlength=1in
\begin{picture}(5.4,2.7)
\put(2.7,0){\epsfig{file=#2.eps, width=2.7in}}
\put(0,0){\epsfig{file=#1.eps, width=2.7in}}
\put(0,2.4){(a)}
\put(2.7,2.4){(b)}
\end{picture}}

\journal{Physics Letters}

\begin{document}
\hfill\parbox{8cm}{\raggedleft TIFR/TH/03-19 \\ hep-ph/0311254 \\
}
\begin{frontmatter}
\title{Resonant slepton production at the LHC in models with an 
  ultralight gravitino}

\author[LAPP]{B.C.  Allanach}, \author[tata]{Monoranjan Guchait}, 
\author[tata]{K. Sridhar}

\address[LAPP]{LAPTH, 9 Chemin de Bellevue, B.P. 110,
  Annecy-le-Vieux 74951, France}
\address[tata]{Tata Institute of Fundamental  
Research, Homi Bhabha Road, Mumbai 400 005, India}

\begin{abstract}
We examine resonant slepton production at the LHC with gravitinos in the final
state. The slepton undergoes gauge decay into a
neutralino and 
a lepton, the neutralino decays into a photon and a gravitino. By measuring
the transverse masses of the $\gamma {\tilde G}$ and the $l \gamma {\tilde G}$ 
subsystems it
is possible to accurately reconstruct both the slepton and the neutralino
masses. 
In some regions of parameter space the slepton decays directly into a lepton
and gravitino, giving an identical experimental topology to $W$
production ($l \mett$). 
We present the novel matrix element squared for lepton-gravitino production.
A peak in the tail of the lepton-missing momentum transverse mass distribution
of the $W$ provides a 
signature for the process and an accurate measurement of the slepton mass. We
display the search reach for the LHC and 300 fb$^{-1}$ of integrated
luminosity. 
\end{abstract}
\end{frontmatter}

\section{Introduction}
This letter is devoted to the study of the signals at the Large Hadron 
Collider (LHC) due to a supersymmetric 
generalisation of the Standard Model (SM) which 
(a) violates $R$-parity, and (b) has an ultra-light gravitino in its 
spectrum. 
Despite the negative results that experimental searches for supersymmetric 
particles have yielded so far, the philosophy underlying these searches
has been rather exclusive. The so-called minimal supergravity
inspired $R$-conserving model of minimal supersymmetry or the constrained
minimal supersymmetric standard model (CMSSM), as it is sometimes called,
has been elevated to the status of a paradigm in the quest
for supersymmetry at colliders. 
The adherence to this model
to the exclusion of a myriad of other possibilities is not desirable. 
This 
is motivation enough to consider other models which
relax some of the assumptions made in the CMSSM. 

The quest to depart from the narrow confines of the CMSSM inexorably leads
one to question the assumption of $R$-parity conservation as well as seeking
alternatives to the mass spectrum of supersymmetric (SUSY)
 particles predicted by the minimal
supergravity models. Of these two paths, the second 
is certainly more interesting, connected as it is with the fundamental 
issue of SUSY breaking. Of particular interest is the question of the lightest
superparticle (LSP) because the decay patterns of the heavier particles into
the LSP decide several of the collider search strategies for supersymmetry.
In the preferred models, the LSP (most often the neutralino) is assumed to be
in the mass range just above the reach of present experiments i.e. of 
$O(100)$~GeV in mass. All the other sparticles, are by definition, heavier
than the LSP (the next-to lightest we denote the NLSP). A dramatic alternative
is to have an ultralight gravitino as 
the LSP. The gravitino, in some models of supersymmetry breaking such as
gauge-mediated supersymmetry breaking \cite{gmsb}, can be as light as 
$10^{-3}$~eV: other sparticles then decay into the gravitino 
and this can significantly alter the strategies for 
supersymmetric particle searches at colliders. The other scenario wherein 
the LSP is not stable is that of $R$-parity violation where
the decays of the neutralino yield final-states with jets, leptons or
missing-$E_T$. But $R$-parity violation has another interesting implication,
{\em ie} the sparticles can be produced singly. If this single sparticle
production process proceeds via the $s$-channel, then the cross-section
is resonantly enhanced and may be significant even for small values
of the $R$-parity violating coupling. 

An anomaly in the CDF experiment in the production 
rate of lepton-photon-$\mett$ in $p{\bar p}$ collisions at $\sqrt{s}= 1.8 
~{\rm TeV}$ was observed using 86.34 pb$^{-1}$ of Tevatron 1994-95 data
\cite{CDF}. 
While the number of events expected from the SM is
$7.6\pm 0.7$, the experimentally measured number corresponded to 16. 
Moreover, 11 of these events involved muons (with 4.2 $\pm$ 0.5 expected)
and 5 electrons (with 3.4 $\pm$ 0.3 expected). These anomalous events
can be simply understood \cite{us} in terms of a supersymmetric model with 
the following features: (1) the model is $R$-parity violating with an 
$L$-violating $\lambda'_{211}$ coupling, and (2) the supersymmetric 
spectrum includes an ultra-light gravitino of mass $\sim10^{-3}$ eV. 
The resonant production of a smuon via the $R$-violating coupling,
its decay into neutralino and a muon and, finally, the decay of the
neutralino into a gravitino and a photon 
leads to the $\mu \gamma \mett$ final state studied in the CDF experiment.
The range of smuon and neutralino masses relevant to the explanation of these 
anomalous observations of the CDF experiment is such that most of
this range will be explored at the Run II of the Tevatron. 
In the event
that this signal is not seen at Run II it will rule out the model at
the lower end of the neutralino and smuon masses. 

For heavier
smuon and neutralino masses (above 250 GeV, roughly), the aforementioned
Run I signal would be a statistical fluke and will probably disappear in Run II
data. In that case, the LHC can be expected to 
discover and measure the sparticle masses. In this paper, we perform a 
study of the ability of the LHC to perform these two tasks, identifying 
the sensitive observables.


\section{The model}
\begin{figure}
\begin{center}
\begin{picture}(400,170)
\ArrowLine(60,10)(120,50)
\ArrowLine(120,50)(60,90)
\DashLine(120,50)(180,50){5}
\ArrowLine(180,50)(240,10)
\ArrowLine(240,90)(180,50)
\ArrowLine(300,60)(240,90)
\Photon(240,90)(300,130){5}{4}
\put(65,5){$q$}
\put(65,90){$\bar q'$}
\put(245,10){$l$}
\put(210,80){$\chi_0$}
\put(150,65){$\tilde l$}
\put(305,55){$\tilde G$}
\put(305,125){$\gamma$}
\end{picture}
\caption{Feynman diagram of resonant slepton production followed by 
neutralino decay.}
\label{feyn}
\end{center}
\end{figure}
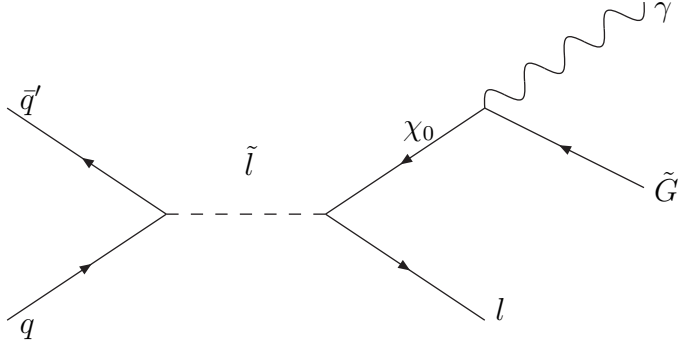

We assume throughout this paper that of all the $R$-violating couplings,
only one lepton-number violating coupling is significantly large while
the others are too small to have observable consequences. The
coupling appears as
\beq
 \lambda'_{ijk} L_i Q_j D_k \subset W
\eeq
in the superpotential (where $i,j,k$ are family indices, gauge indices have
been suppressed and $L_i$, $Q_j$, $D_k$ denote the left-handed lepton doublet,
the left-handed quark doublet and the charge-conjugated right-handed down
quark superfields respectively).
This coupling
allows for the production of a slepton from the initial 
state of quarks. The slepton thus produced could decay again via 
the $R$-violating coupling into
two quarks. However, if the $R$-violating coupling is small, the existence 
of an ultralight gravitino of a mass $\sim {\mathcal O}(10^{-3})$~eV
drastically  
alters the decay mode of the slepton. The slepton overwhelmingly decays into 
a lepton and a (bino-dominated) neutralino, with the latter decaying 
into a photon and a gravitino resulting in a $l\gamma\mett$ final-state. 
The Feynman diagram for the process is shown in Fig.~\ref{feyn}. 
Since $m_{\tilde \nu} < m_{{\tilde l}_L}$ we should also
expect signals from sneutrino production.
In the case of sneutrino production, the process $q \bar{q}' \rightarrow 
{\tilde \nu} \rightarrow \gamma {\tilde G} \nu$ gives rise to 
a $\gamma\mett$ final-state. We will not study this final state
in this paper, since the background depends crucially on cosmic ray events
which are difficult to estimate.
In our model, the only other light sparticles are the neutralino (which
is lighter than the slepton) and the 
ultra-light gravitino (which is as light as $10^{-3}$ eV). We enforce
degeneracy between the first two generations in order to avoid flavour
changing neutral currents. Other sparticles do not play a role in this
analysis, and are set to be arbitrarily heavy.

Such a light
gravitino materialises naturally in models of gauge-mediated supersymmetry
breaking (GMSB)\cite{gmsb}. However, in minimal GMSB models
the chargino and the second-lightest neutralino are not much heavier than 
the neutralino. This is certainly not desirable for our considerations 
because it predicts large jets$+ \gamma +
\mett$ rates which are not seen by experiments\footnote{This 
observation has been made earlier in the literature \cite{baer}
in the context of the GMSB-based explanation \cite{kane} of the
$ee \gamma \gamma \mett$ CDF event~\cite{abe}.}. 
The question that naturally follows is how we can deviate from
the minimal versions of GMSB models and obtain the pattern of
supersymmetric masses that we require. It is interesting
to note that such a mass spectrum can arise in an alternate model of GMSB 
which is obtained from compactifying 11-dimensional M-theory on a 7-manifold 
of G2 holonomy~\cite{witten}. However, the detailed low-energy predictions
of this model have not been worked out so for our analysis we simply work
with the low-energy model described above.

\section{Simulation Results}

For our study of the process shown in Fig.~\ref{feyn} at the LHC ($pp$ 
collisions at the $\sqrt s = 14$ TeV), we have chosen to work with the
following default set of model parameters (unless indicated otherwise):
\begin{itemize}
\item
Gravitino mass, $m_{\tilde G}=10^{-3}$~eV,
\item 
$R$-violating coupling $\lambda'\equiv\lambda'_{211}=0.01$,
\item
${\rm tan}\beta = 10$,
\item
sparticle masses $(m_{\chi_1^0}, m_{\tilde l})=(120\gev,200\gev)$ or $(200\gev,500\gev)$ 
GeV (``low mass'' and ``high mass'' scenarios) respectively.
\end{itemize}
The choice of using $\lambda'_{211}$ rather than some other flavour
combination is arbitrary.
Our results can be easily generalised to other $R$-violating
couplings  
and we will comment upon this below. 
Note that the chosen value of $\lambda'_{211}$ is much smaller than
the bound coming from  
$R_\pi = \Gamma (\pi \rightarrow e \nu) / (\pi \rightarrow \mu \nu)$
\cite{bgh}: $\lambda'_{211}< 0.059 \times \frac{m_{\tilde{d_R}}}{100~{\rm GeV}}$ 
\cite{rparrev} even for a squark mass of 100 GeV. However, since the
squark mass is arbitrarily large in our model, this constraint is not relevant
to our analysis. The $R$-violating
decay of the slepton is possible but constrained, in principle, by the
Tevatron di-jet data \cite{cdfjets} which exclude a $\sigma . B> 1.3 \times 
10^4$ pb at 95\%~C.L. for a resonance mass of 200 GeV. However, in practice
this does not provide a restrictive bound upon our scenario as long as 
the $R$-violating coupling is sufficiently small, $< \mathcal{O}(1)$. Moreover,
the di-jet bound is not restrictive because it suffers from a huge QCD 
background. By restricting $\lambda'_{211}$ and the gravitino mass to be
small, we avoid  
significant rates for the possible $R$-violating decays of $\chi_1^0 
\rightarrow \mu jj$ or $\chi_1^0 \rightarrow \nu jj$. $\chi_1^0 \rightarrow
\gamma {\tilde G}$ is the dominant channel.

We use the {\small \tt ISASUSY} part of the {\small \tt ISAJET7.58}
package~\cite{isajet} to generate the spectrum, branching ratios and 
decays of the sparticles. For an example of parameters, we choose 
(in the notation used by ref.~\cite{isajet}) $\tan \beta=10$ and
$A_{t,\tau,b}=0$. $\mu$ together with the flavour-diagonal soft 
supersymmetry breaking parameters are set to be very large. We 
emphasise that this is a representative point in the supersymmetric 
parameter space and not a special choice. 

\subsection{Slepton decays to lepton, photon and gravitino}

The signal has been simulated
using {\small \tt HERWIG6.4}~\cite{herwig}. The $W \gamma$ SM background to the
$\mu \gamma \mett$ and $\gamma \mett$ channels have been simulated 
using {\small \tt PYTHIA}~\cite{Sjostrand:2003wg}. Because the background took
a restrictively long time to calculate, 10 times {\em less} background
luminosity was 
simulated than signal luminosity. 
The background has then been scaled up by a factor
of 10 for all results presented here, but it should be borne in mind that 
the statistical fluctuations within it are bigger than will be expected.
For our simulations of the signal and the 
background we have used only selection cuts of $E_T> 25$ GeV on the
transverse energies of the muon and the photon. The same cut of 25
GeV is also used for $\mett$. We have used the following cuts
on the rapidity of the photon and the muon: $|\eta_{\gamma,\mu}|<3$.
There is an isolation cut between the photon and other hard objects $o$ in the
event of $\sqrt{(\eta_\gamma - \eta_o)^2 + (\phi_\gamma - \phi_o)^2}>0.7$.
Since the signal is hadronically quiet, we veto events with jets
reconstructed with $E_T>30$ GeV and $\eta_j<4$. Initial and final state
radiation effects, as well as fragmentation effects are included in the
background simulation.

\begin{figure}
\begin{center}
\threegraphs{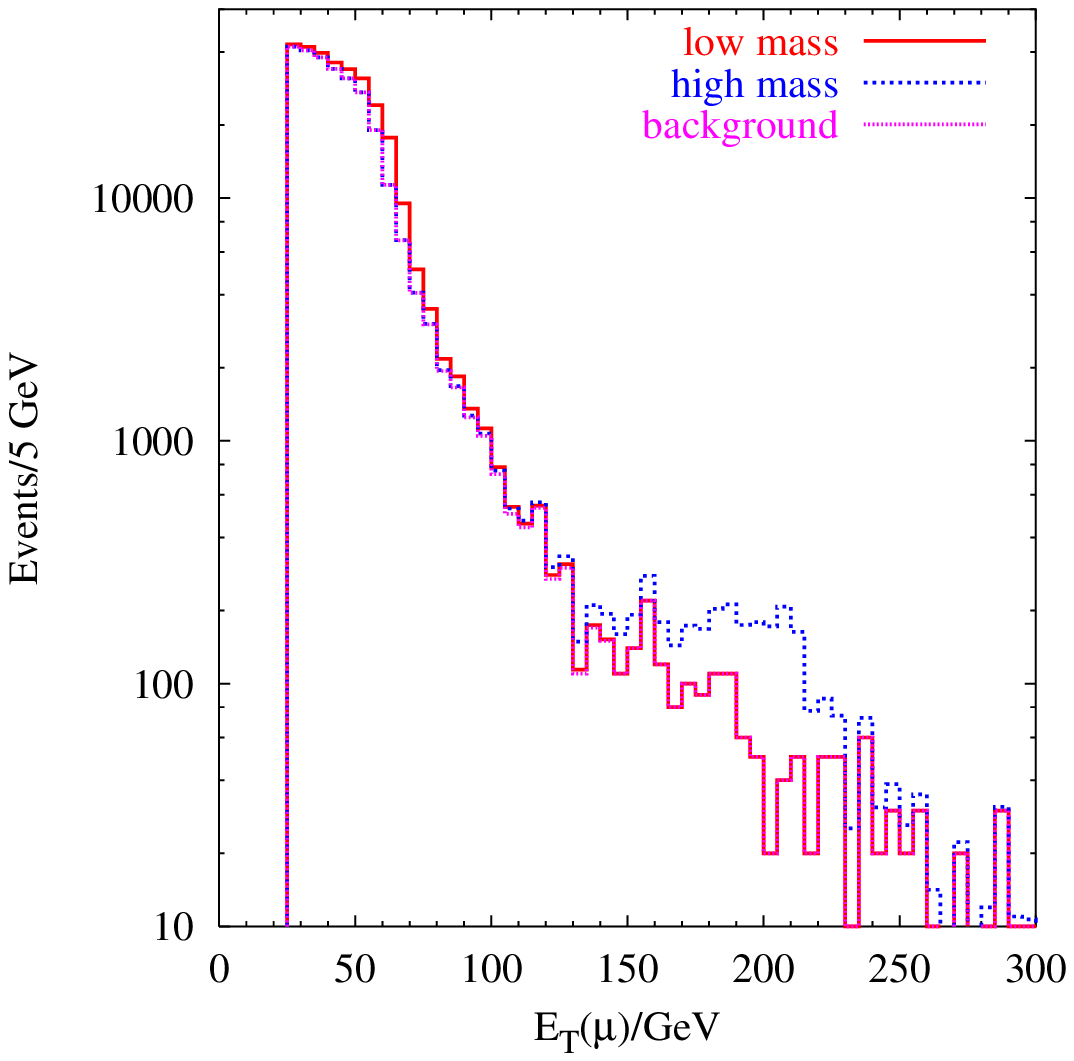}{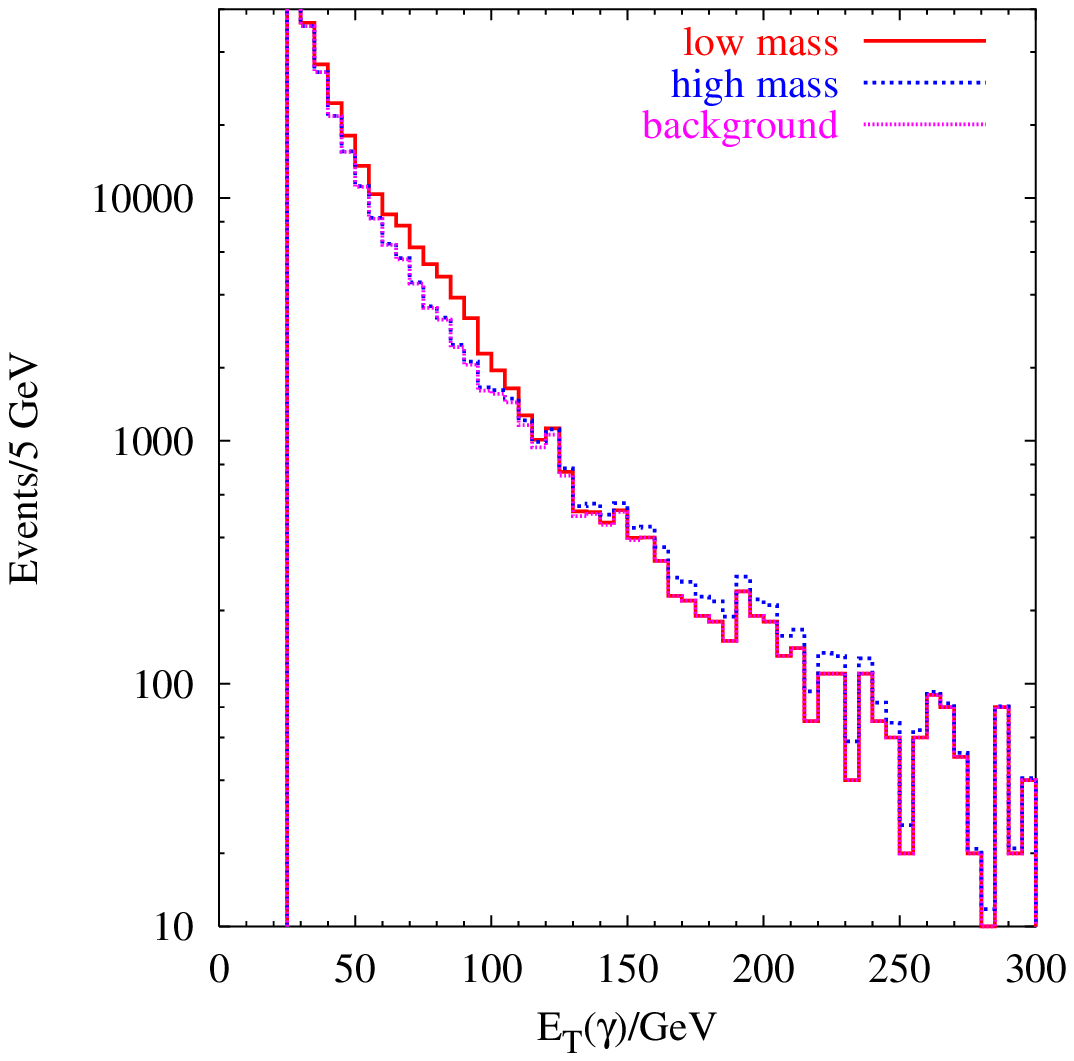}{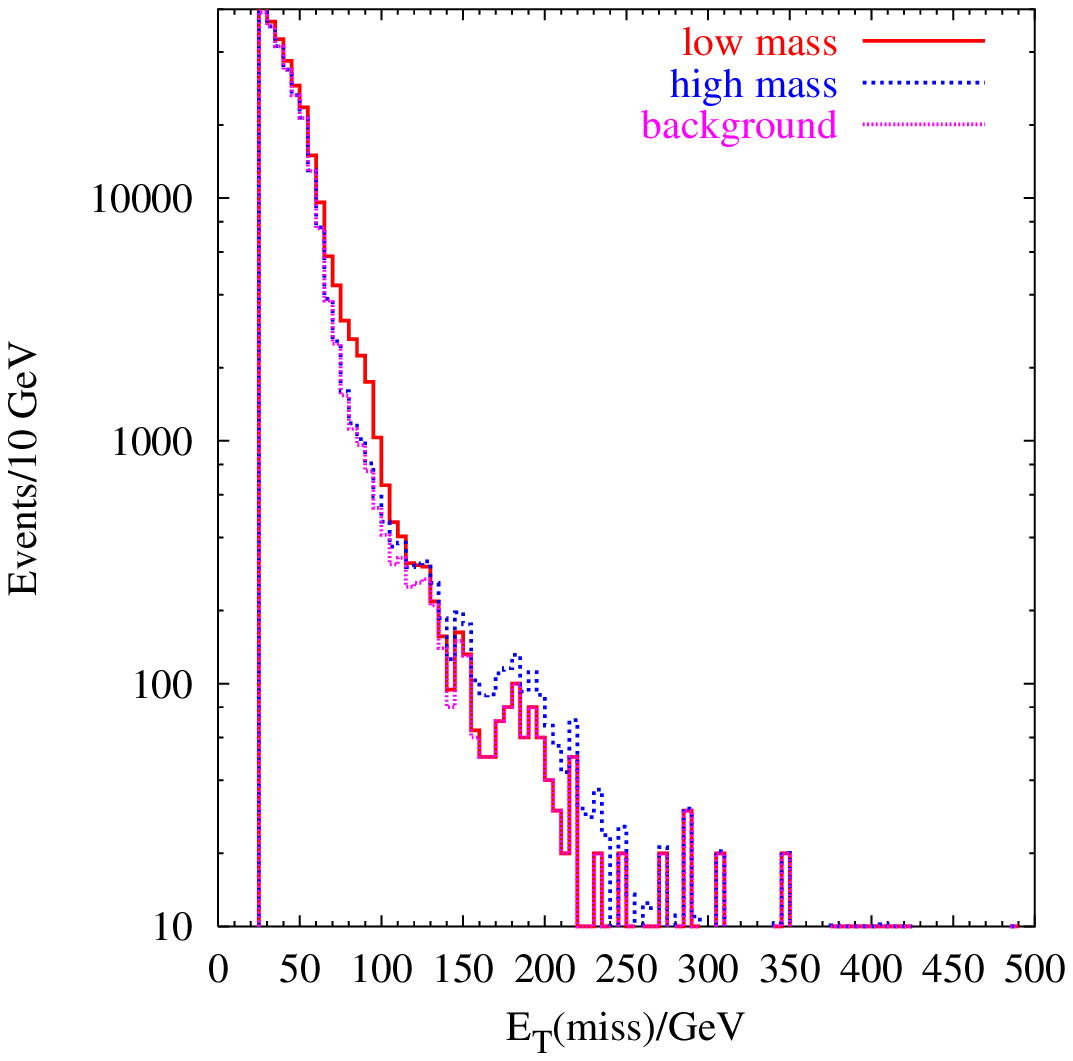}
 \caption{$E_T$ distributions of (a) muons, (b) photons
   and (c) missing-$E_T$ for $\mu
   \gamma \mett$ events that pass the cuts. 300 fb$^{-1}$ integrated luminosity
   at the LHC is assumed.
   The purple (lighter) histograms display $W \gamma$ SM background, the red
   (darker) 
   histograms show the signal plus background for $(m_{\chi_1^0}, m_{\tilde l})=(120\gev,200\gev)$, 
 whereas the blue (dotted) histograms display the signal plus background
   distributions for 
$(m_{\chi_1^0}, m_{\tilde l})=(200\gev,500\gev)$.}
\label{etdist}
\end{center}
\end{figure}
The $E_T$ distributions of $\gamma, \mu$ and $\mett$ in the simulated events
are shown in Figs.~\ref{etdist} (a)-(c). It would be very difficult to claim a
signal for the low or high mass scenarios based on these
distributions, given uncertainties involved in the background
calculation. However, the figures illustrate the fact that the signal to
background ratio could be improved by tightening the $E_T$ cuts, to 50 GeV for
example. We leave the optimisation of cuts to a future, more detailed study.

\begin{figure}
\begin{center}
\unitlength=1in
\begin{picture}(5.4,2.7)
\put(2.7,0){\epsfig{file=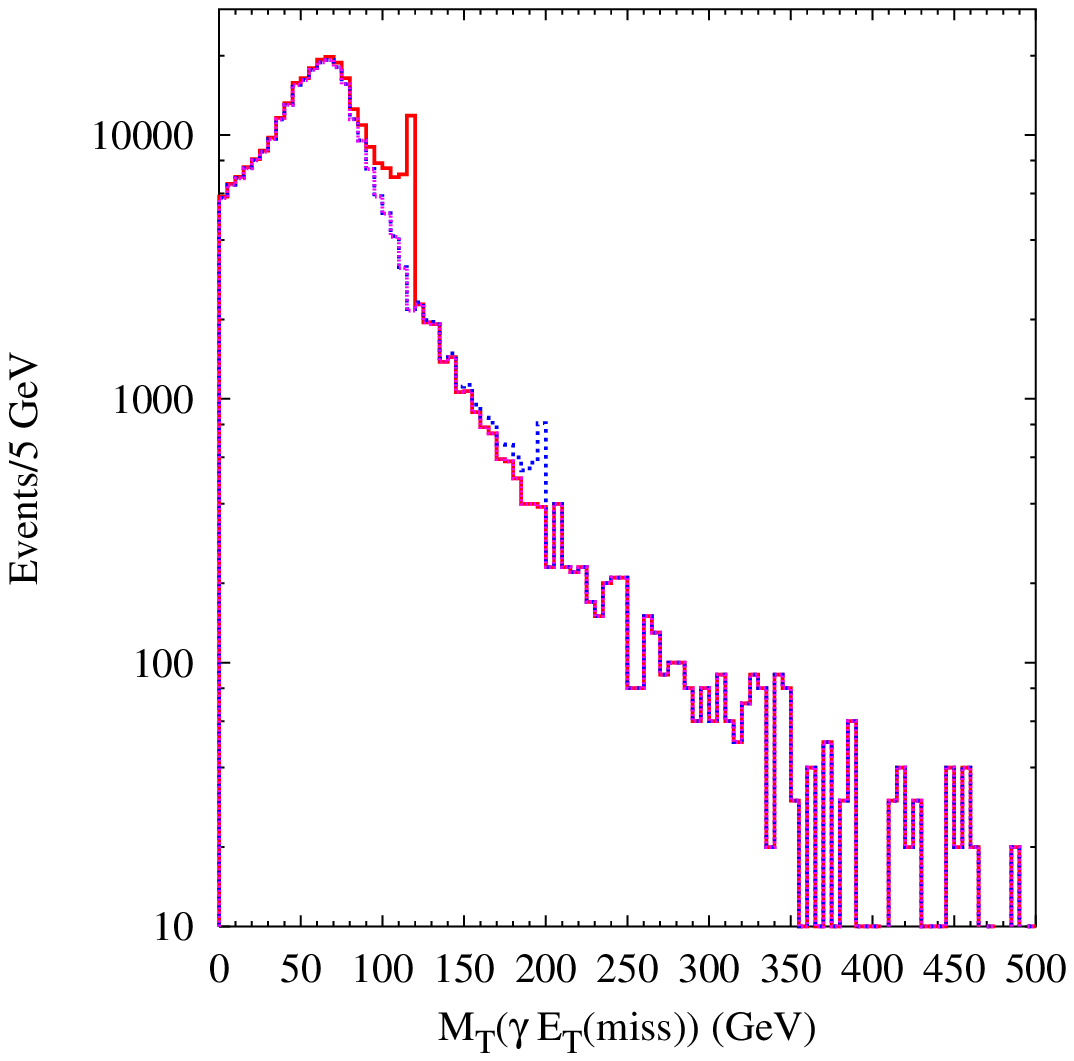, width=2.7in}}
\put(0,0){\epsfig{file=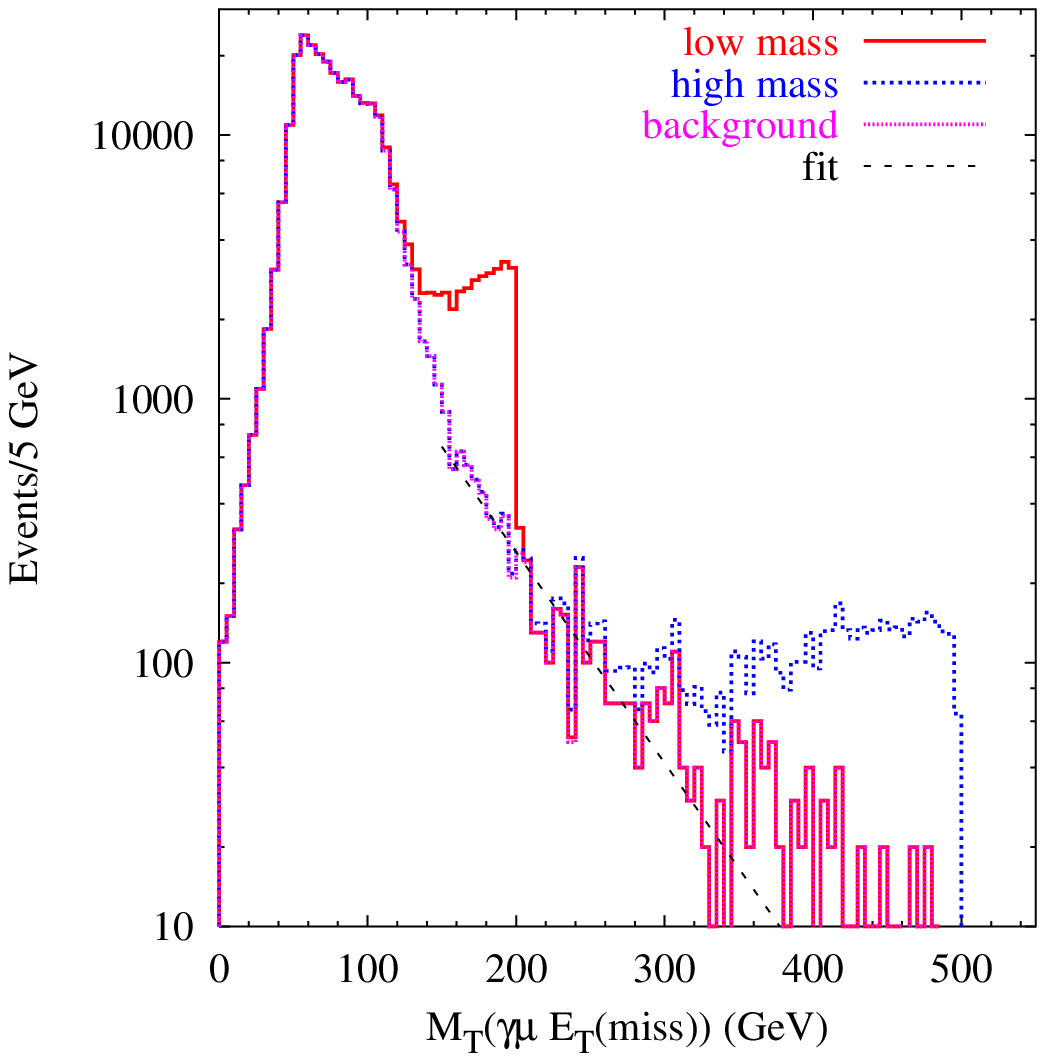, width=2.7in}}
\put(4.05,1.3){\epsfig{file=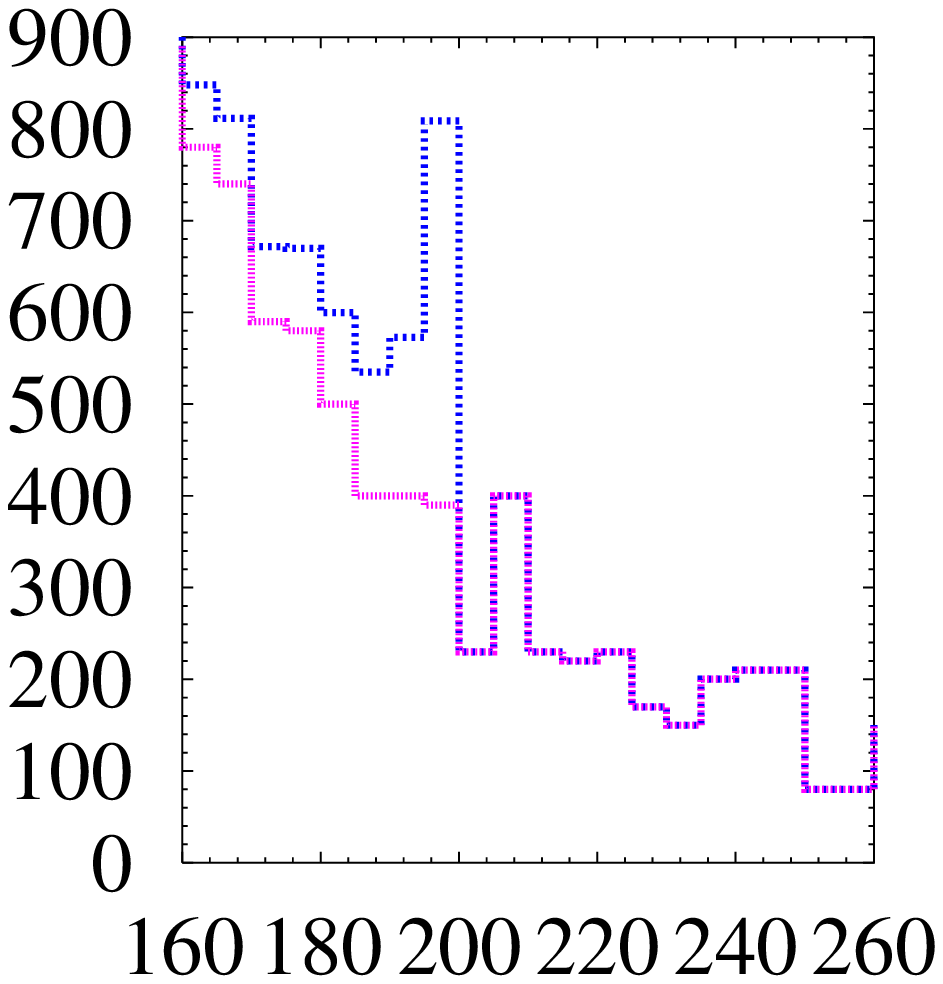, width=1.35in}}
\put(0,2.4){(a)}
\put(2.7,2.4){(b)}
\end{picture}
 \caption{$M_T$ distributions of (a) $\mu \gamma \mett$, (b) 
  $\gamma \mett$ for
  300 fb$^{-1}$ integrated luminosity  at the LHC.
   The purple (lighter) histograms display $W \gamma$ SM background, the red
  (darker) 
   histograms show signal plus background for $(m_{\chi_1^0}, m_{\tilde
  l})=(120\gev,200\gev)$,  
 whereas the blue (dotted) histograms display the signal plus background distributions for
$(m_{\chi_1^0}, m_{\tilde l})=(200\gev,500\gev)$.
In (a), the dashed black line displays a log-linear fit to the background
  distribution for $M_T=150-400$ GeV. In (b), the insert shows a linear scale
  magnification of 
  an area of the plot.}
\label{mtdist}
\end{center}
\end{figure}
The transverse mass distributions of several particles $i$
\beq
M_T = \sqrt{(\sum_i E_T^i)^2 - (\sum_i {\underline p}_T^i)^2}
\eeq
should possess a sharp peak at the mass of the particle which decays 
to $i$ daughters. The $M_T(\mu \gamma \mett)$ and $M_T(\gamma \mett)$
distributions 
are displayed in Figs.~\ref{mtdist}a,b for the simulated high and low
mass points and $W \gamma$ simulated SM background. In Fig.~\ref{mtdist}a,
sharp peaks in the $M_T(\mu \gamma \mett)$ are clearly
visible at values of the smuon mass and will be detected above the SM $W
\gamma$ background. The smuon mass could be accurately measured in this manner.
The background gives a negligible number of events above
$M_T(\mu \gamma \mett)>450$ GeV. 
Fig.~\ref{mtdist}b shows that the signal peaks in $M_T(\gamma \mett)$
(predicted to be at the neutralino mass) should be able to
provide a measurement of the neutralino mass. 

\begin{figure}
\begin{center}
\twographs{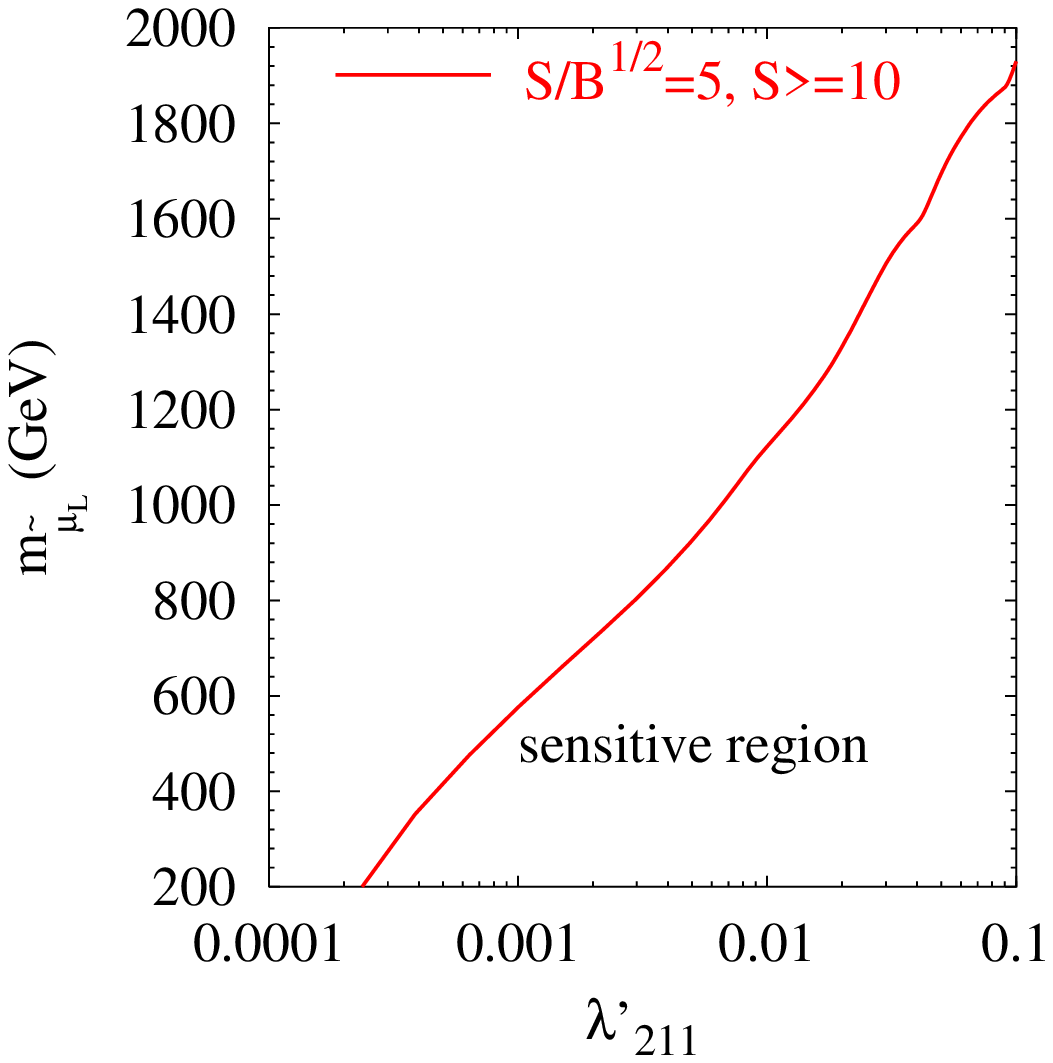}{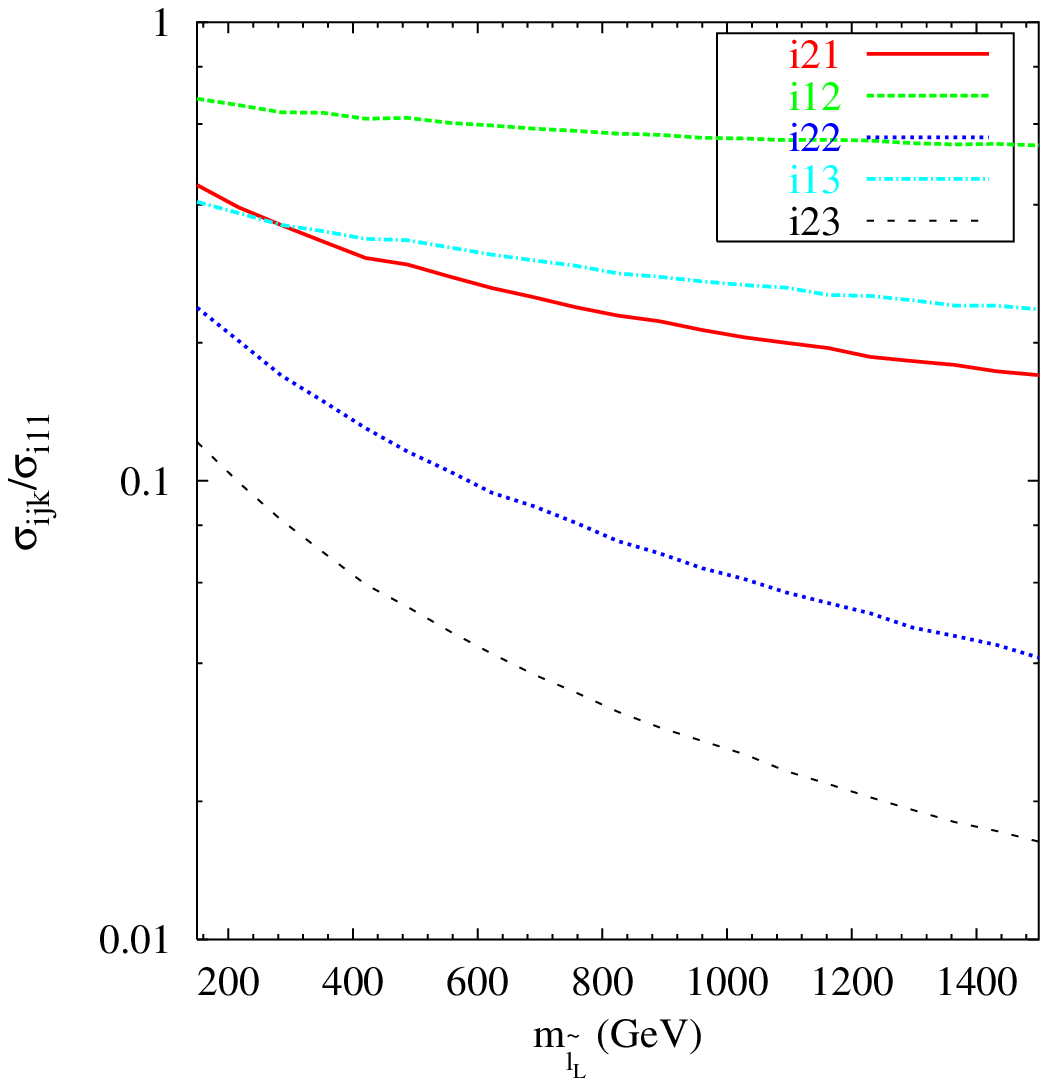}
 \caption{(a) Search reach for the $\mu \gamma \mett$ signal (as defined in the
   text) for
   300 fb$^{-1}$ integrated luminosity  at the LHC. 
   (b) Relative production cross-sections for the
   different flavour choices of $\lambda'_{ijk}$
   couplings as a function of the slepton mass. 
}
\label{search}
\end{center}
\end{figure}
In order to calculate the search reach, we use the signal $S$ in the
4 highest peak bins (covering 20 GeV) of $M_T(\mu \gamma \mett)$. The
background distribution in these four bins is estimated by fitting a simple
function to $M_T(\gamma \mu \mett)$ between 150-400 GeV in
Fig.~\ref{mtdist}a.
We use
\begin{equation}
B = 4 \mbox{exp} \left[a M_T(\mu \gamma \mett) + b\right]
\end{equation}
with purely $\sqrt{B}$ statistical errors, leading
to $a=-0.018\pm0.001$, $b=9.25\pm0.22$. $B$ is displayed in Fig.~\ref{mtdist}a
as the dashed black line.
We define the sensitive region of parameter space to be any region for
which\footnote{The 
  statistical uncertainties
  on fitted $a$ and $b$ parameters make a negligible difference to the final
  numerical results.} $S/\sqrt{B}>5$
and $S\geq10$.
For 300 fb$^{-1}$, the search reach is shown as a function of smuon mass 
and R-parity conserving coupling in Fig.~\ref{search}a for the case of a neutralino NLSP.

If a different R-parity violating (RPV) flavour coupling than $\lambda'_{211}$ was used, 
the analysis would be much the same provided the coupling did not involve taus
or top quarks. 
The main difference would be that, for a given size of RPV
coupling,  the production cross-section would decrease
 because one no longer probes
valence partons in the proton. Fig.~\ref{search}b shows that the suppression
of the cross-section also depends upon the slepton mass, since different 
regions of $x$ are being probed in the parton distributions. For a fixed 
slepton mass, 
the ratio of production cross sections 
$\sigma(\lambda_{ijk}=c) / \sigma(\lambda_{i11}=c)$
is shown.
The parton
distributions of top quarks are not known, so we do not include the
possibility that the RPV coupling involves the top ($\lambda'_{i3j}$). The
figure shows that the production cross-section can decrease 
to less than 10$\%$ of its original value by changing the flavour structure of
the RPV coupling to include heavy quarks. 
The number of produced events for different flavours of RPV coupling other
than
$\lambda'_{211}$ can therefore be determined by multiplying the ratio of cross
sections in Fig.~\ref{search}b with the number of events in
previous plots.
One must then ask how the number of {\em measured} signal events depends upon
the lepton flavour of the RPV coupling used. 
Efficiencies and backgrounds for the $e$ and $\mu$ case will differ
somewhat depending upon the experiment, and so changing the lepton flavour
involved in the RPV coupling 
could quantitatively affect the result. $\tau$ reconstruction efficiencies
are significantly different to those of $e,\mu$ and so there would be a large 
change in number of measured events in that case.

There is no particular need to rely on a neutralino NLSP.
Indeed, recent studies have found sleptons to be lighter than neutralinos in,
for example, 
large regions of RPV mSUGRA space~\cite{Allanach:2003eb}. Later (for a
different final state), we will
consider the 
possibility of a smuon NLSP, and contrast it with the neutralino NLSP case.

\subsection{Slepton decays to gravitino and lepton}

We now turn to the decay ${\tilde l} \rightarrow {\tilde G} l$. We ignore
sneutrino production in this case because it would lead to an invisible final
state.
Neglecting lepton masses, the partial widths for the decays of ${\tilde
  l}$~\cite{gmsb,bgh} are 
\bea
&&\Gamma ({\tilde l} \rightarrow {\tilde G} l) = \frac{m_{\tilde l}^5}{48 \pi
  M_P^2 m_{\tilde G}^2}, \qquad
\Gamma ({\tilde l} \rightarrow q \bar{q}') = \frac{3 {\lambda'}^2
m_{\tilde l}}{16 \pi} \nonumber \\
&& \Gamma ({\tilde l} \rightarrow l \chi_1^0) = |b|^2 \frac{m_{\chi_1^0}}{8
  \pi} \left[
\frac{m_{\chi_1^0}}{m_{\tilde \mu}}
\left(1 - \left(\frac{m_{\chi_1^0}}{m_{\tilde \mu}}\right)^2  \right)^2 \right],
\eea
where $b=e N_{11} + g N_{12} (1/2 - \sin^2 \theta_w) / \cos \theta_w$ and
$N_{ij}$ denote neutralino mixing matrix elements~\cite{peter}.
$\Gamma ({\tilde l} \rightarrow {\tilde G} l)$
thus dominates for large $m_{\tilde l}$ and small $m_{\tilde G}$.

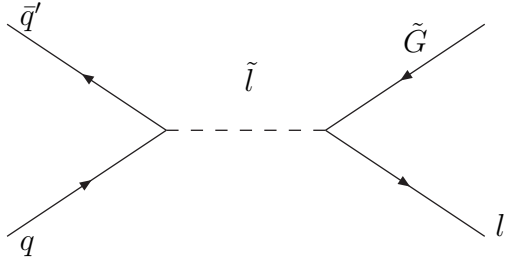
\begin{figure}
\begin{center}
\begin{picture}(250,170)
\ArrowLine(60,10)(120,50)
\ArrowLine(120,50)(60,90)
\DashLine(120,50)(180,50){5}
\ArrowLine(180,50)(240,10)
\ArrowLine(240,90)(180,50)
\put(65,5){$q$}
\put(65,90){$\bar q'$}
\put(245,10){$l$}
\put(210,80){$\tilde G$}
\put(150,65){$\tilde l$}
\end{picture}
\caption{Feynman diagram of resonant slepton production followed by decay into
a lepton and gravitino}
\label{feyn2}
\end{center}
\end{figure}
The Feynman diagram of the process in question is displayed in
Fig.~\ref{feyn2} and we calculate the spin and colour averaged matrix element
squared to be: 
\beq
| \bar{M} |^2 = 
\frac{\lambda'^2 m_{\tilde l}^4}{36 M_P^2 m_{\tilde G}^2}
\frac{s^2}
{(s-m_{\tilde l})^2 + \Gamma^2 m_{\tilde l}^2},
\eeq
where $s$ is the centre-of-mass energy of the colliding $q \bar{q}'$ system,
$M_P= 2.4\times 10^{18}$ GeV is the reduced Planck mass and $\Gamma$ is the
total decay width of the slepton.

This new process is not simulated in {\tt HERWIG}, so we perform the phase
space integration by the method of Monte-Carlo integration ourselves. For this
reason, the simulation was not of generated events and is therefore not
subject to associated statistical errors.
We have use cuts 
on the muon and missing transverse energy identical
to the $\gamma \mu \mett$ analysis, i.e. 
$\mett,E_T^\mu>25$ GeV and $|\eta_\mu|<3$.

\begin{figure}
\begin{center}
\twographs{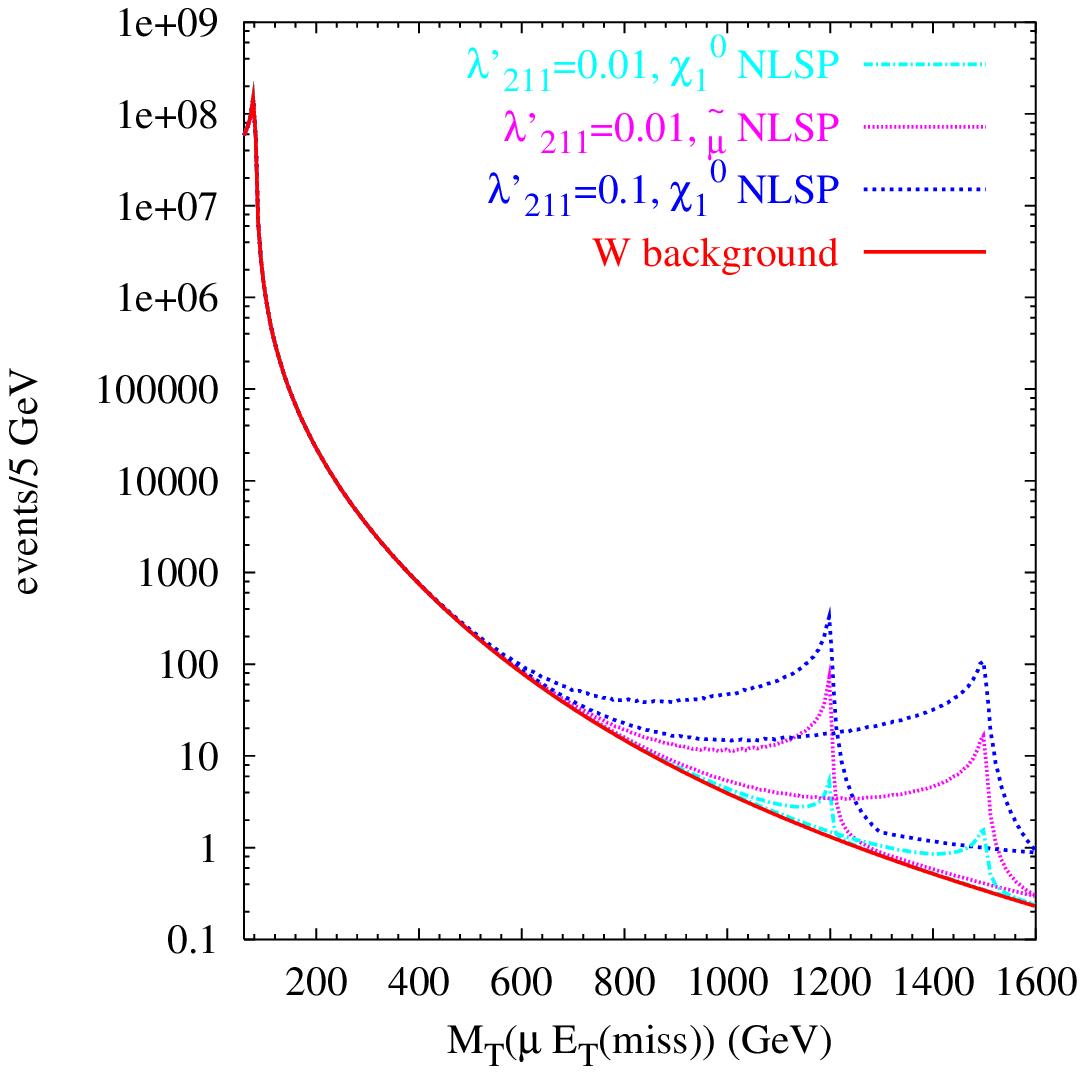}{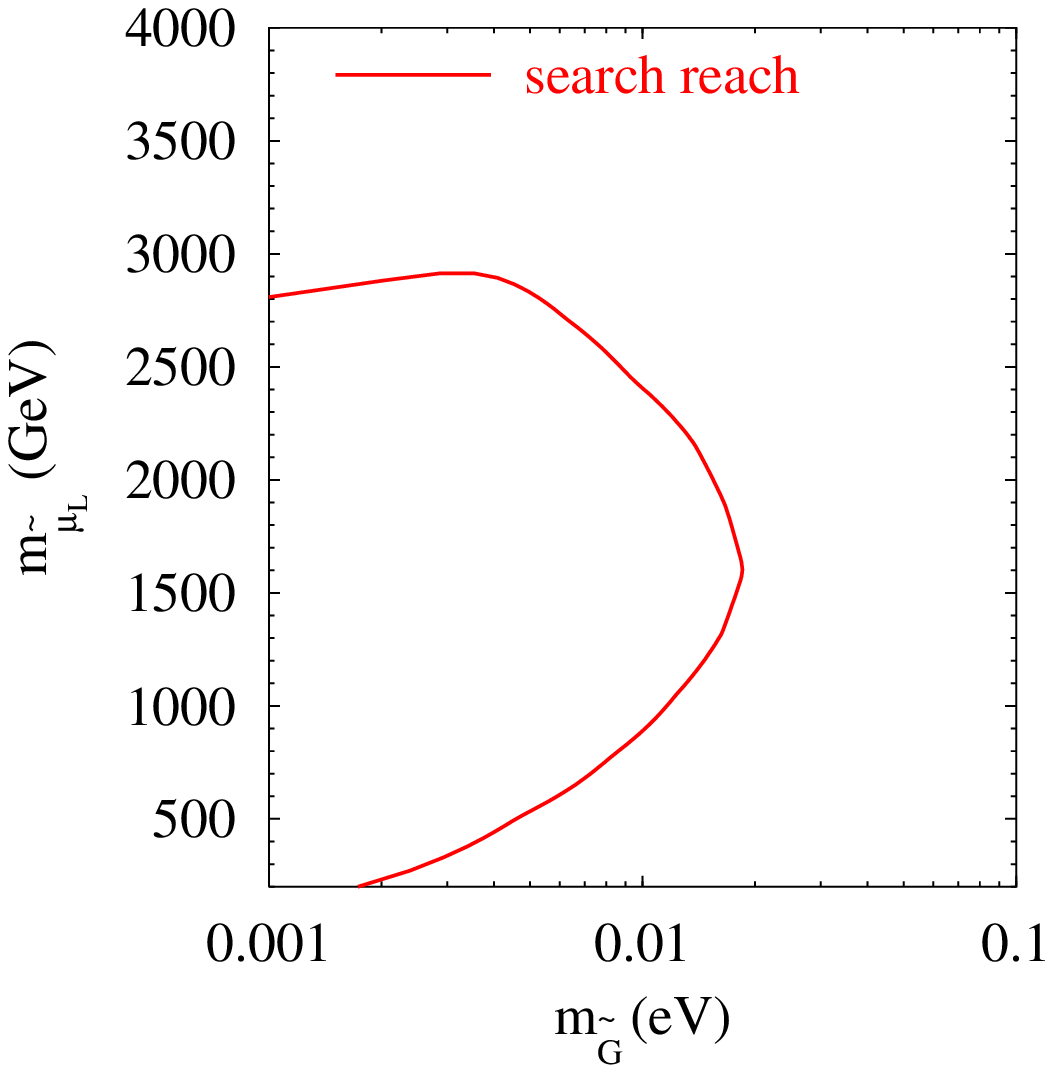}
 \caption{(a) $M_T$ distribution of the $\mu \tilde G$ final state
  for two different values of the smuon mass and 300 fb$^{-1}$ integrated
  luminosity  at the LHC. The left(right)-most three peaks are for smuon masses
  of 1.2 and 1.5 TeV respectively.
  (b) Search reach (as defined in the text) for the $\mu \tilde G$ final
  state, a {\em neutralino} NLSP
  and 300 fb$^{-1}$ integrated
  luminosity  at the LHC and $\lambda'_{211}=0.1$. 
  The search reach is contained to the left of the
  curve.}
\label{new}
\end{center}
\end{figure}
The $M_T(\mu \mett)$ distribution is displayed in Fig.~\ref{new}a for the SM
$W$ background plus signal in two cases: $m_{\tilde \mu}=1.2$ TeV, $1.5$
TeV. If we use the default value of $\lambda'_{211}=0.01$ and the neutralino
is the NLSP, we obtain the
smaller peaks which would not be prominent enough to claim a $5\sigma$
discovery. If the smuon is the NLSP, the decay mode to neutralinos is not
open, leading to a higher branching ratio for ${\tilde \mu} \rightarrow
{\tilde G \mu}$. The magenta (middle height) 
curves in Fig.~\ref{new}a are evidence for this, 
and should be observable even for the small value of $\lambda'_{211}=0.01$.
The signal cross section scales as ${\lambda'_{211}}^2$, and for
$\lambda'_{211}=0.1$ we obtain the one-hundred times
larger peaks, either of 
which would be easy to detect on top of the SM $W$ background,
even for a neutralino NLSP.

In this subsection, we define the search reach by the criteria that for $S$
signal and $B$ 
background events, $S/\sqrt{B}\geq 5$ and $S\geq 10$ in the 5 GeV
signal peak bin of $M_T(\mu \mett)$. These constraints lead to the
expected search reach for 
300  fb$^{-1}$ luminosity shown in Fig.~\ref{new}b for a neutralino NLSP. For
smuon masses that 
are too low, the huge background swamps the signal in the peak bin. For smuon
masses that are too high, the signal production cross-section is too low. 
The search region is shown as the region to the left of the curve in 
Fig.~\ref{new}b.

\begin{figure}
\begin{center}
\twographs{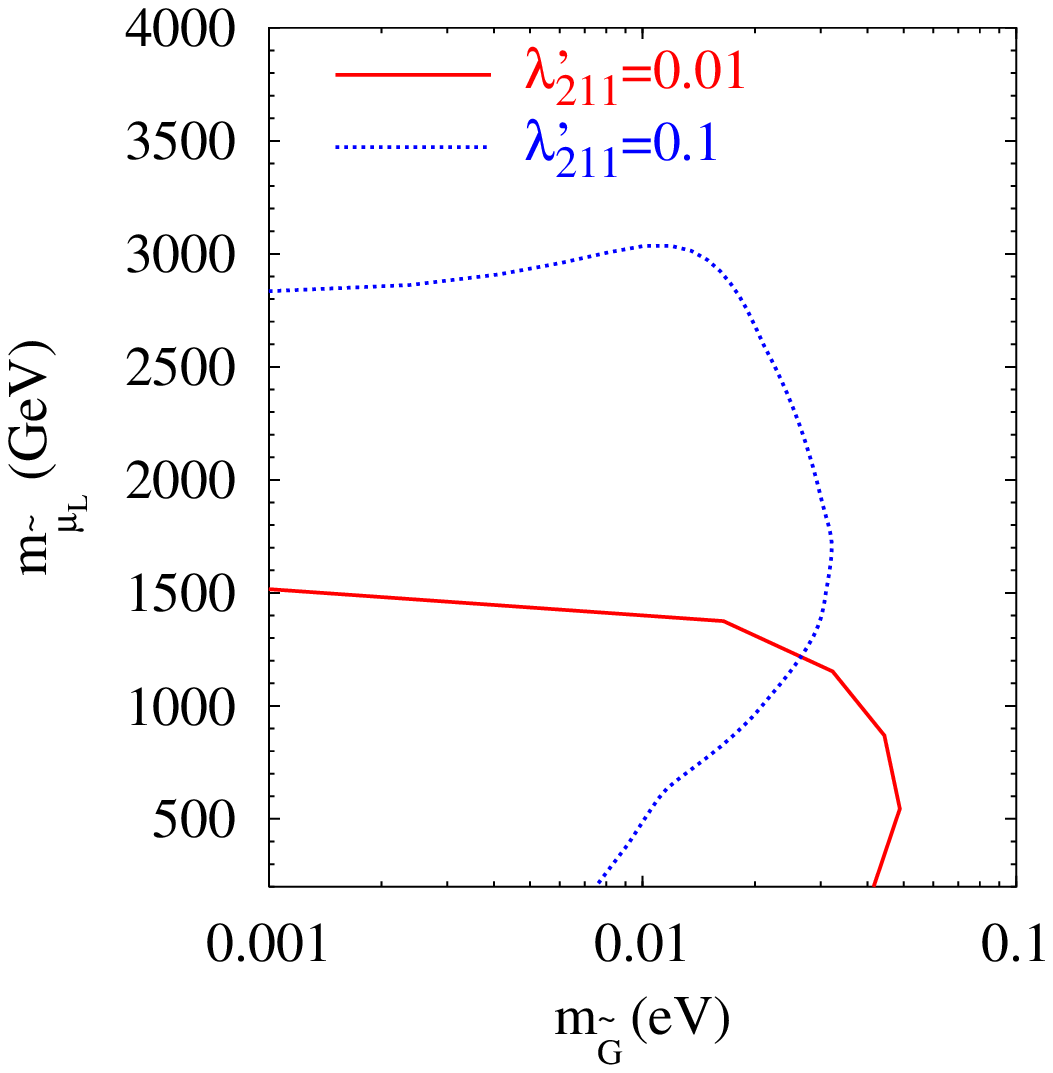}{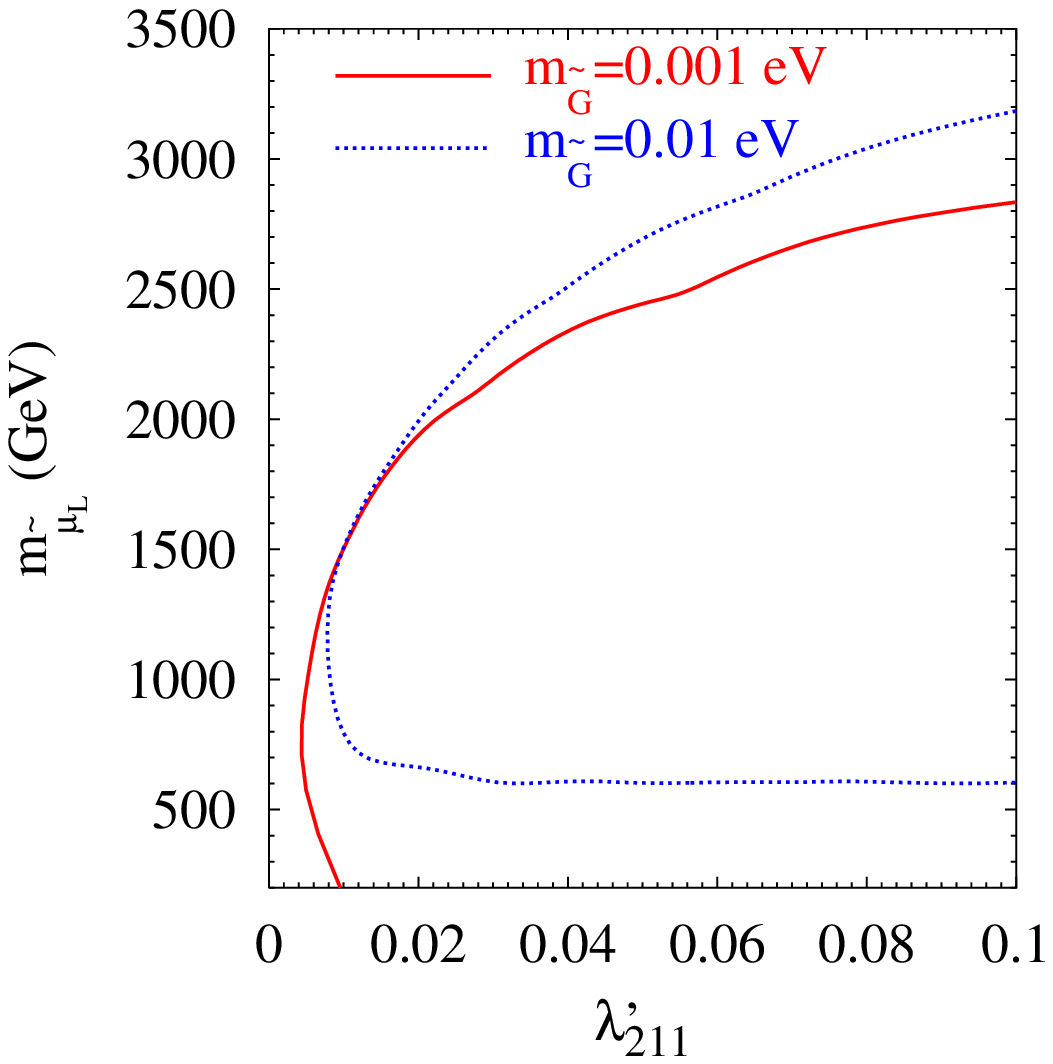}
 \caption{Region of sensitivity to the $M_T(\mu {\tilde G})$ peak for 300
 fb$^{-1}$ at the LHC for a {\em smuon} NLSP. The region of sensitivity is
(a) to the left hand side of the curves, (b) to the right hand side of the
 curves.}
\label{new2}
\end{center}
\end{figure}
As explained above, it is easier to detect the $\mu {\tilde G}$ final state if
the smuon is the NLSP, because the relevant branching ratio is larger. Thus we
expect a larger region of sensitivity. The search reach in two different
planes of parameter space are displayed in Fig.~\ref{new2} for the case of a
smuon NLSP.
For $\lambda'_{211}=0.1$, the region of sensitivity is modestly
larger than the neutralino NLSP case, as can be seen by comparing the blue
curve in Fig.~\ref{new2} with the one in Fig.~\ref{new}b. However, there is no
possibility searching for a peak with $\lambda'_{211}=0.01$ with a neutralino
NLSP, whereas a smuon NLSP is still detectable up to $m_{\tilde \mu} \approx
1500$ GeV, as displayed in Fig.~\ref{new2}a.

\section{Conclusions}

We have provided a basic first study of the search for resonant slepton
production at the LHC with ultra light gravitinos in the final
state\footnote{We note that resonant slepton production at the LHC has been
  studied before without decays into gravitinos in Ref.~\protect\cite{hhh}, for example.}.
We have not performed a detailed detector simulation, with associated
mis-tagged backgrounds, $\mett$ resolution and detection efficiencies. 
We started with a simple and loose set of cuts, and provided distributions
which indicate how signal to background ratio can be improved.
Specific cases of RPV coupling and other parameters are taken in order to be
concrete, but the analysis applies over a wide range of parameter space.

The current paper shows which are the sensitive variables (various
$M_T$ distributions are the most useful) and displays the rough search reach
one can expect. Luckily, signal peaks in $M_T$ distributions mean that 
one does not have to know the backgrounds very well. The backgrounds can be
measured and fitted away from the peaks, whereas the position of the peak
provides an accurate measurement of the slepton mass.
Although detector efficiencies and non-physics backgrounds
will erode the search reach somewhat, it could be possible to flavour
subtract some backgrounds. SM background processes from $W \rightarrow l
\nu$ are lepton-flavour universal. On the other hand, if one RPV coupling is
dominant over others, the signal only contributes to the production of one
lepton flavour. Thus, subtracting the number of electron-tagged events from
the number of muon-tagged events would greatly improve the signal to
background ratio. The only pitfall of this approach is that it is possible
that the parameter space is such that one produces similar numbers of  
selectrons and smuons and therefore one could inadvertently cancel 
the signal. Although this only happens in a very limited region of parameter
space, one cannot {\em a priori} rule this pessimistic scenario out and so we
have not included it in our analysis.
The process $q \bar{q}' \rightarrow {\tilde l} \rightarrow {\tilde G} l$ has
not been studied before, and we presented the matrix element squared together
with the search reach and $M_T$ distributions.

We have concentrated on LHC studies, but of course the same ${\tilde \mu}
\rightarrow \mu {\tilde G}$
production analysis could be
applied to the Tevatron, with lighter smuons than those considered
here. The important indicator would still be a peak in the $M_T(\mu)$ tail,
above $M_W$. Instead of smuon production, we could instead have considered
squark production through a baryon-number violating coupling $1/2 \lambda^{\prime\prime}_{ijk}
U_i D_j D_k$ in the superpotential. We would then have signals of $j \gamma
\mett$ in the neutralino NLSP case\footnote{$j$ denotes a hard jet.} and 
$j \mett$ (``monojet'' signature). The sensitivity of the Tevatron and LHC to
these processes is presumably less than in the slepton-production case, but
remains to be calculated. In that case, the same $M_T$ variables will be
appropriate for detection, but some of the cuts may need to be harder in order
to beat down a larger background. 

\section*{Acknowledgements}
This work was conceived and commenced at the Les Houches workshop {\em Physics
  at Tev colliders}. We would like to thank Giacomo Polesello for discussions
and helpful comments  during the Les Houches workshop.
K Sridhar would like to thank LAPTH for hospitality
  offered while part of this work was carried out.

\end{document}